\documentclass[aps,prb,twocolumn,a4paper]{revtex4}
\usepackage{amsmath}
\begin{document}
\title{Site-averaging in the integral equation theory of interaction site
models of macromolecular fluids: An exact approach}

\author{V. Krakoviack\footnote{Present address: Laboratoire de Chimie,
\'Ecole Normale Sup\'e\-rieure de Lyon, 46, All\'ee d'Italie, 69364 Lyon
Cedex 07, France}}

\affiliation{Department of Chemistry, University of Cambridge,
  Lensfield Road, Cambridge CB2 1EW, United Kingdom}

\begin{abstract}
A simple "trick" is proposed, which allows to perform exactly the
site-averaging procedure required when developing integral equation
theories of interaction site models of macromolecular fluids. It shows
that no approximation is involved when the number of Ornstein-Zernike
equations coupling the site-site correlation functions is reduced to
one. Its potential practical interest for future theoretical
developments is illustrated with a rederivation of the so-called
molecular closures.
\end{abstract}

\maketitle

In the recent years, the application of liquid state integral equation
methods \cite{simpleliquids} to soft matter systems allowed
significant advances in our understanding of their properties. One of
the most prominent developments in this field is the formulation of
the Polymer Reference Interaction Site Model (PRISM), extending to
macromolecular systems the Reference Interaction Site Model (RISM) of
Chandler and Andersen.\cite{chandler} Mostly applied to polymeric
fluids,\cite{PRISM} it has recently been considered to study fluids of
platelet-like colloids.\cite{harcoshan01el}

It is usually derived from the RISM theory as follows. Consider a
one-component fluid with molecules made of $N$ identical sites,
i.e. the interaction potentials between any pair of sites on different
molecules are the same. These sites are in general non-equivalent
(except those related by obvious symmetries) and, at the pair level,
the structure of the fluid has thus to be characterized by a matrix of
$N\times N$ site-site total correlation functions $h_{ij}(r)$, where
indices $i$ and $j$ refer to types of sites on different
molecules. The RISM approach relies on the generalized
Ornstein-Zernike matrix equation
\begin{equation}\label{RISM}
h_{ij}(r)=\sum_{k,l=1}^N \omega_{ik}*c_{kl}*[\omega_{lj}
+\rho\,h_{lj}](r),
\end{equation}
where $\rho$ is the molecular number density and $*$ denotes
convolution integrals. $c_{ij}(r)$ is the direct correlation function
between sites of types $i$ and $j$ on different molecules, and
$\omega_{ij}(r)$ the intramolecular distribution function of sites of
types $i$ and $j$. To solve Eq.~\eqref{RISM}, it has to be
supplemented by some appropriate closure. In the case of hard-sphere
sites of diameter $d$, it can be the Percus-Yevick approximation
$h_{ij}(r)=-1$ if $r<d$, $c_{ij}(r)=0$ if $r>d$.
\cite{simpleliquids,chaand72jcp}

PRISM is obtained by making the simplifying assumption that all $N$
sites are equivalent. This is true for ring molecules for instance,
but it amounts to a neglect of end effects for linear molecules or of
edge effects for platelets. Then, using $h_{ij}(r)\simeq h(r)$ and
$c_{ij}(r)\simeq c(r)$ for all $i,j$'s, one finds that
Eq.~\eqref{RISM} reduces to the single scalar equation
\begin{equation}\label{PRISM}
h(r)=\omega*c*[\omega+\rho\,N\,h](r),
\end{equation}
with
\begin{equation}
\omega(r)=\frac{1}{N}\sum_{i,j=1}^N \omega_{ij}(r).
\end{equation}
$\omega(r)$ is the total intramolecular distribution function.  Using
a perturbation theory approach, Curro and Schweizer \cite{cursch87jcp}
have shown that the best choices for $h(r)$ and $c(r)$ are the site
averaged values,
\begin{equation}\label{distrib}
h(r)=\frac{1}{N^2}\sum_{i,j=1}^N h_{ij}(r)
\end{equation}
and similarly for $c(r)$, thus demonstrating that PRISM involves a
pre-averaging of chain end (or platelet edge) effects and evidencing
the practical relevance of the theory, since the doubly summed
quantity of Eq.~\eqref{distrib}, to which it gives access, appears
naturally in the calculation of the scattering properties or of the
thermodynamics of the system.

In the present Note, I want to put forward a ``trick'' for the
derivation of PRISM from RISM equations, showing that the reduction of
the $N\times N$ matrix equations of the latter to the scalar equation
of the former can in fact be done without invoking any approximation.

The basic idea of this derivation is that, if all sites on a molecule
are identical in term of their interaction potential, then there is a
permutation symmetry between them. To take this symmetry into account,
one can artificially decompose each molecule into two objects: a
backbone made of $N$ sites, labelled $1$ to $N$, in which are embedded
the geometry and the intramolecular interactions of the original
molecule, and a set of $N$ identical ``atoms'', labelled $\hat{1}$ to
$\hat{N}$, carrying the intermolecular interactions and occupying at
random the backbone sites such that all permutations of $N$ atoms have
equal probability.  Clearly, this delocalisation of the interaction
sites does not correspond to any real physical process and is only
introduced for formal consideration.

One can write RISM equations for the mobile atoms, which have the
usual form \eqref{RISM}, now with hatted indices. These simply are
formal definitions of the direct correlation functions
$c_{\hat{i}\hat{j}}(r)$, applicable to any molecule within a site-site
description.

In this system, all interaction sites are by construction
equivalent. Thus, for all $\hat{i}$ and $\hat{j}$, one has
$h_{\hat{i}\hat{j}}(r)=h'(r)$, $c_{\hat{i}\hat{j}}(r)=c'(r)$, and
$\sum_{\hat{j}=\hat{1}}^{\hat{N}}\omega_{\hat{i}\hat{j}}(r)=
\omega'(r)$. In addition, it is obvious that the total distribution
functions of the present model coincide with the site-averaged
quantities obtained within the conventional RISM approach,
i.e. $h'(r)=h(r)$ and $\omega'(r)=\omega(r)$. It immediately follows
from Eq.~\eqref{RISM} with hatted indices that Eq.~\eqref{PRISM}, with
$c$ replaced by $c'$, is exactly obeyed. Note that nothing can be said
\emph{a priori} on the relation between $c$ and $c'$ and in particular
there is no reason to think they should be equal.

So, what did we learn ? From a practical point of view, not much:
within the present approach, the reduction of RISM to PRISM has been
made a tautology and, owing to the odd nature of the backbone plus
delocalized atoms system, the final situation is somewhat equivalent
to simply taking Eq.~\eqref{PRISM} as a formal definition of the total
direct correlation function. The problem can be considered as only
being displaced to the question of the non-trivial relation between
the $c_{ij}$'s of the usual RISM approach and the
$c_{\hat{i}\hat{j}}$'s of the present fictitious system. However, the
fundamental point remains: the reduction from $N^2$ to $1$ of the
number of Ornstein-Zernike equations does not involve any
approximation and, at variance with the usual claim, nothing like end
or edge effects is lost when going from the RISM level of description
to the simpler PRISM one. As usual with integral equation theories, if
PRISM, compared to a full RISM treatment, can appear inadequate in
some case, it is only because of our inability to guess a correct
expression for the direct correlation function.

More interesting is probably the fact that the present picture could
provide new insights on the PRISM formalism and thus lead to
improvements of the theory. To illustrate this point, I propose a
derivation of the so-called molecular closures \cite{yetsch92jcp}
developed for models where the intermolecular site-site interaction
consists of a hard core part of diameter $d$ plus a soft tail
potential.

Starting with the classical RISM approach, $c_{ij}(r)$ for $r>d$ is
often given the interpretation of a renormalized intermolecular
site-site interaction potential. In the backbone plus delocalized
atoms system, it can thus be considered as an effective interaction
between backbone sites. It is effective because the backbone sites do
not directly interact and the interaction is actually mediated by the
delocalized particles. Consider a configuration in which particle
$\hat{k}$ on one molecule and particle $\hat{l}$ on another molecule
occupy respectively the backbone sites $i$ and $j$. Using the simplest
superposition approximation, the contribution to $c_{ij}(r)$ of two
particles, $\hat{m}$ on the same molecule as $\hat{k}$ and $\hat{n}$
on the same as $\hat{l}$, can be approximated as
$\omega_{\hat{k}\hat{m}}*c_{\hat{m}\hat{n}}*\omega_{\hat{n}\hat{l}}(r)$.
Summing over all mobile particles and averaging over all their
permutations, one eventually finds $c_{ij}(r)\simeq
\omega*c'*\omega(r)$ for $r>d$. \cite{note} Then, using a so-called
atomic closure for $c'(r)$, Eq.~\eqref{PRISM} with $c$ replaced by
$c'$, and the impenetrability constraint $h(r)=-1$ if $r<d$, we are
left with the same equations as those proposed in
Ref.~\onlinecite{yetsch92jcp}, but with a somewhat different physical
content. This points to possible improvements of the molecular
closures: indeed, it has for instance been shown that in some cases,
simply convoluting by $\omega$ is not the best way to give an account
of the macromolecular connectivity. \cite{krahanlou01el}

Finally, like in PRISM, equivalence of all sites on a molecule is
often assumed in the framework of the density functional theory, in
order to derive equations for macromolecular fluids from theories of
heteronuclear site-site fluids.\cite{DFT} One could thus think of
using the present ``trick'' in this context. However, since it is
already known that the excess free-energy functional can exactly be
expressed in terms of the total monomer density\cite{woo91jcp} (which
can be thought of as a site-averaged quantity for inhomogeneous
systems), here again the proposed approach would probably not lead to
significant practical progress, but only help clarifying the formal
relation between the site-specific and site-averaged levels of
description.

The author acknowledges support from the EPSRC under grant number
GR/M88839 and thanks J.-P. Hansen, A. A. Louis, R. Allen, and
D. G. Rowan for useful discussions.


\begin{thebibliography}{99}
\bibitem{simpleliquids} See e.g. J. P. Hansen and I. R. McDonald,
\textit{Theory of Simple Liquids, 2nd Ed.} (Academic Press, London,
1986).

\bibitem{chandler} For a review, see D. Chandler, in \textit{Studies
in Statistical Mechanics}, edited by E.~Montroll and J. L.~Lebowitz
(North-Holland, Amsterdam, 1982), vol.~8, p.~275.

\bibitem{PRISM} For a review, see K. S. Schweizer and J. G. Curro,
Adv. Chem. Phys. \textbf{98}, 1 (1997).

\bibitem{harcoshan01el} L. Harnau, D. Costa, and J.-P. Hansen,
Europhys. Lett. \textbf{53}, 729 (2001).

\bibitem{chaand72jcp} D. Chandler and H. C. Andersen,
J. Chem. Phys. \textbf{57}, 1930 (1972).

\bibitem{cursch87jcp} J. G. Curro and K. S. Schweizer,
J. Chem. Phys. \textbf{87}, 1842 (1987).

\bibitem{yetsch92jcp} A. Yethiraj and K. S. Schweizer,
J. Chem. Phys. \textbf{97}, 5927 (1992); K. S. Schweizer and
A. Yethiraj, J. Chem. Phys. \textbf{98}, 9053 (1993).

\bibitem{note} Note that $c_{ij}(r)$ is found independent of $i$ and
$j$ in this approximation.

\bibitem{krahanlou01el} V. Krakoviack, J.-P. Hansen, and A. A. Louis,
Europhys. Lett. \textbf{58}, 53 (2002).

\bibitem{DFT} See for instance J. P. Donley, J. G . Curro, and
J. D. McCoy, J. Chem. Phys. \textbf{101}, 3205 (1994); A. Yethiraj,
J. Chem. Phys. \textbf{109}, 3269 (1998); A. Yethiraj, H. Fynewever,
and C.-Y. Shew, J. Chem. Phys. \textbf{114}, 4323 (2001); T. Sumi and
F. Hirata, J. Chem. Phys. \textbf{118}, 2431 (2003).

\bibitem{woo91jcp} C. E. Woodward, J. Chem. Phys. \textbf{94}, 3183
(1991).

\end{thebibliography}
\end{document}